# Bipolar electric-field switching of perpendicular magnetic tunnel junctions through voltage-controlled exchange coupling

Delin Zhang,[1]* Mukund Bapna,[2]* Wei Jiang,[1]* Duarte Pereira de Sousa,[1]* Yu-Ching Liao,[3] Zhengyang Zhao,[1] Yang Lv,[1] Protyush Sahu,[1] Deyuan Lyu,[1] Azad Naeemi,[3] Tony Low,[1†] Sara A Majetich,[2†] Jian-Ping Wang[1†]

[1] Department of Electrical and Computer Engineering, University of Minnesota, Minneapolis, Minnesota 55455, USA.
[2] Department of Physics, Carnegie Mellon University, Pittsburgh, Pennsylvania 15213, USA.
[3] School of Electrical and Computer Engineering, Georgia Institute of Technology, Atlanta, GA 30332 USA

*These authors contributed equally to this work.
†Corresponding authors. Email: jpwang@umn.edu (J.P.W.), sara@cmu.edu (S.A.M.) and tlow@umn.edu (T.L.)

**Perpendicular magnetic tunnel junctions (p-MTJs) switched utilizing bipolar electric fields have extensive applications in energy-efficient memory and logic devices. Voltage-controlled magnetic anisotropy linearly lowers the energy barrier of ferromagnetic layer via electric field effect and efficiently switches p-MTJs only with a unipolar behavior. Here we demonstrate a bipolar electric field effect switching of 100-nm p-MTJs with a synthetic antiferromagnetic free layer through voltage-controlled exchange coupling (VCEC). The switching current density, ~$1.1\times10^5$ A/cm$^2$, is one order of magnitude lower than that of the best-reported spin-transfer torque devices. Theoretical results suggest that electric field induces a ferromagnetic-antiferromagnetic exchange coupling transition of the synthetic antiferromagnetic free layer and generates a field-like interlayer exchange coupling torque, which cause the bidirectional magnetization switching of p-MTJs. A preliminary benchmarking simulation estimates that VCEC dissipates an order of magnitude lower writing energy compared to spin-transfer torque at the 15-nm technology node. These results could eliminate the major obstacle in the development of spin memory devices beyond their embedded applications.**

Spin memory devices such as perpendicular magnetic tunnel junctions (p-MTJs) can be switched through the spin-transfer torque (STT) effect to store data (*1*). However, its high energy consumption remains a key challenge for many applications because of the charge current flowing through p-MTJs. Using an electric-field (E-field) to control the magnetization switching has been considered as a promising method and has been intensively studied for the applications on ultralow power memory and logic devices (*2-8*). Recently, many efforts have been focused on the voltage-controlled magnetic anisotropy (VCMA) effect, where a voltage (E-field) induces electron accumulation or depletion at the ferromagnetic (FM)/MgO interface, modifying the magnetic properties (e.g. magnetic anisotropy), and realizing the magnetization switching of FM layers (*9-18*). However, the VCMA effect is generally an odd function of applied voltage, i.e., the positive (negative) voltage can only reduce (enhance) the magnetic anisotropy of ferromagnetic (FM) layers. This phenomenon results in unipolar E-field switching of the magnetization of p-MTJs with the assistance of an external magnetic field ($H_{ext}$) or the STT effect (*14-16*).

A synthetic antiferromagnetic (SAF) structure is composed of two FM layers which couple together through a non-magnetic (NM) layer (as a spacer) via a Ruderman-Kittel-Kasuya-Yosida (RRKY) exchange interaction (a type of interlayer exchange coupling (IEC)). This IEC can be manipulated by changing the thickness of the FM layer or the spacer layer (*19,20*). Recently, Yang et al. (*21*) and Newhouse-Illige et al. (*22*) reported the experimental E-field-induced IEC modulation in a SAF structure and MTJ stack, respectively, through a magneto-ionic effect, in which the E-field changes the effective thickness of the spacer or the FM layer. The operation speed of such a device needs to be improved further since it relies on the diffusion of ions. On the other hand, Bruno's theory indicates that an energetic preference for exchange coupling configuration can be selectively manipulated by the modulation of the spin reflectivity at the interfaces (23). Based on Bruno's theory, You, et al. (24) and Fechner, et al. (25) theoretically found that the voltage can modulate the spin reflectivity (spin-up and spin-down) at the interfaces in the SAF structure, inducing a sign change in the IEC. This concept sheds light on a new possibility for bipolar E-field controlled magnetization switching of p-MTJs with a SAF free layer, but until now experimental demonstration has not yet been reported.

Here, by taking advantage of the physics of Bruno's theory, we designed and fabricated sub-100 nm p-MTJ devices with a FePd-$Co_{20}Fe_{60}B_{20}$ (CFB) SAF free layer and demonstrated bipolar E-field switching via voltage-controlled exchange coupling (VCEC) without an assistance of $H_{ext}$. We found that the E-field with bipolar polarity can induce the linear-in-voltage field-like torque originated from IEC of the SAF free layer, inducing the bidirectional magnetization switching of the p-MTJ devices. This avoids the damage of the p-MTJ memory devices on the operation with a low/high unipolar E-field and eliminates the major obstacle in the development of spin memory devices beyond their embedded applications.

FePd is a promising candidate for the development of the ultrahigh-density and ultralow energy spintronic devices because of a large bulk perpendicular magnetic anisotropy (PMA) ($K_u >$ 10 Merg/cm$^3$) and low damping constant ($\alpha \sim 0.006$) (26) as well as the sensitive magnetic property change under E-field (4,27). To demonstrate such VCEC switching of p-MTJs, we designed a FePd SAF p-MTJ stack, as shown in Fig. 1A, where an SAF free layer structure (FePd/Ru/FePd/Ta/CFB) is utilized in this work. Two FePd PMA layers couple together via a Ru spacer and then couple with a CFB layer through a Ta spacer (28) (see Fig. S1). The FePd SAF p-MTJ stack was annealed at 350 °C and then characterized by the high-resolution cross-sectional scanning transmission electron microscopy. The stack shows a highly textured layer structure, as shown in Fig. 1B. To investigate its spin transport properties, the FePd SAF p-MTJ stack was patterned into nano-sized MTJ pillars (see Fig. 1C). Subsequently, the magnetization switching of these FePd SAF p-MTJ nanopillars was tested using a conductive atomic force microscope (C-AFM) setup without applying $H_{ext}$ (see Fig. S2, and Methods and Materials). In all the C-AFM measurements, the tip was grounded and $V_{bias}$ (positive or negative) was applied at the bottom electrode. Therefore, the positive (negative) $V_{bias}$ corresponds to positive (negative) E-field (as shown in Fig. S2B). Figure 1D shows the current vs. bias voltage ($I$-$V_{bais}$) curves of the 100-nm FePd SAF p-MTJ pillars with a 2.0-nm MgO tunnel barrier. $V_{bias}$ is swept from -0.6 V to +0.6 V. A bidirectional magnetization switching is observed in the FePd SAF p-MTJs (see the insets of Fig. 1D) with bipolar $V_{bias}$ (parallel-to-antiparallel state at $V_{bias}$ ~ -0.45 V and antiparallel-to-parallel state at $V_{bias}$ ~ +0.48). The switching current density, $J_c$, is calculated to be ~ $1.1 \times 10^5$ A/cm$^2$ based on the switching $V_{bias}$, which is one order of magnitude lower than that of the best reported STT devices. Furthermore, to evaluate the contribution of the STT effect, we calculated the effective field ($H_{STT}$) originated from the STT effect based on our experimental data, as shown in Fig. S3. The maximum calculated $H_{STT}$ ~ 5 Oe suggests that the STT effect could be excluded for this bipolar switching (see Section S2 in Supporting Information (SI)).

Figure 2A shows this VCEC switching mechanism using a simplified model. The schematic MTJ stack is composed of a top FM$_3$ reference layer and a bottom FM$_2$/NM/FM$_1$ SAF free layer. We first set the initial magnetization state of the FM$_2$ layer in antiparallel to the magnetization of the FM$_3$ reference layer (the FM$_1$ and FM$_2$ layers form an antiferromagnetic (AFM) exchange coupling). When applying a negative E-field, the exchange coupling between the FM$_1$ and FM$_2$ layers changes from AFM to FM exchange coupling, and the magnetization of the FM$_2$ layer switches from up to down, which results in the antiparallel configuration between the FM$_2$ and FM$_3$ layers. The p-MTJ is then in a high resistance state. When a positive E-field is applied, the exchange coupling between the FM$_1$ and FM$_2$ layers is changed back to the AFM exchange coupling, and the magnetization of the FM$_2$ layer points up. In this case, the FM$_2$ and FM$_3$ layers are in the parallel configuration, and the p-MTJ has a low resistance. A qualitative understanding of physical origin of the VCEC in this p-MTJ structure could be as follows. The electron wave functions penetrate the MgO tunnel barrier with the penetration length modulated by E-field (as shown in Fig. 2B). This modulation of the penetration length effectively modulates the reflection phases, which are different for the two-electron spins (spin up and spin down) due to their different band offsets. Since the exchange coupling energy strongly depends on these reflectivities, the voltage modulation can then induce an FM-AFM exchange coupling transition, which will be further elaborated in a later section of this manuscript.

To demonstrate the robustness of magnetization switching in FePd SAF p-MTJs, the current vs. time trace was measured while applying voltage pulses on the same p-MTJ device without $H_{ext}$. The results are plotted in Figs. 3A and 3B. During the testing, ±0.85 V voltage pulses were applied to *write*, and a small *read* voltage of -0.1 V was utilized to monitor the changes in current values in-between the *write* pulses, as shown in Fig. 3A. It is clearly observed that the positive/negative voltage pulses can lead to the low- and high-resistance states toggling back and forth, demonstrating bidirectional magnetization switching via VCEC, as depicted in Fig. 3B. In our devices, the IEC strength of the SAF free layer is about 0.056 erg/cm$^2$ (~ 1×10$^5$ erg/cm$^3$ normalized by the thickness of the free layer) without E-field. This value is smaller than the interfacial PMA ($K_u$ =3×10$^5$ erg/cm$^3$) of the CFB layer (See Fig. S4D) (*29*). When applying the negative E-field, the AFM to FM exchange coupling transition happens due to a dramatically enhancement of the IEC strength of the FePd-CFB SAF layer and the decrease of the magnetic anisotropy of the CFB layer, as shown in Fig. 4B and Fig. S4D. When removing negative E-field, the interfacial PMA of the CoFeB layer goes back to the initial value. At the same time, the SAF free layer will prefer to recover back to the initial state (AFM exchange coupling). However, because the interfacial PMA of the CoFeB layer is larger than the interlayer exchange coupling energy between the FePd and CoFeB layers, therefore the SAF free layer will keep the FM state when removing E-field. For the positive E-field case, the SAF free layer always keeps the AFM state. Furthermore, to evaluate the E-field effect in FePd SAF p-MTJs, the multiple minor resistance vs. magnetic field (*R-H*) loops were measured in 100-nm MTJ devices at different $V_{bias}$ (see Section S1 in SI) (*29*). Figure S4A plots the *R-H* loops with different $V_{bias}$. It is seen that the center of the *R-H* loop with $V_{bias} \sim$ -0.1 V shifts to the left side of zero field, indicating that the CFB layer in the bottom free layer suffers the influence of the stray field ($H_{stray}$) from the rest of the FM layers. Through micromagnetic simulations, we found that $H_{stray}$ on the CFB layer is ~ 400 Oe, which matches the center shift of the *R-H* loop with $V_{bias} \sim$ -0.1 V (see Fig. S5). Meanwhile, we found that the center of the *R-H* loops at $V_{bias} \sim$ +0.75 V and -0.75 V also shifts to the left side of the zero field ~ 290 Oe and ~ 680 Oe, respectively, as plotted in Fig. S4B. Because the very small STT effect in this MTJ device cannot induce such large center shifts (as shown in Fig. S3), this phenomenon implies that the voltage changes the exchange coupling of the SAF structure (*30*). In our p-MTJ devices, the exchange coupling between the free layer and reference layer through a MgO tunnel barrier can be ignored due to the thick MgO tunnel barrier (*31*). Therefore, the exchange coupling change is only contributed from the SAF free layer. The center position of a minor *R-H* loop reveals the exchange coupling field ($H_{ex}$) of a SAF structure, which represents FM coupling or AFM coupling of the SAF structure (*29*). If setting the center of *R-H* loop with $V_{bias} \sim$ -0.1 V as a reference point, we found that the center of the *R-H* loop with $V_{bias}$ ~ +0.75 V shifts to the right and the center of the *R-H* loop with $V_{bias}$ ~ -0.75 V shifts to the left (labeled with the arrow in Fig. S4A), implying the FM-AFM exchange coupling transition in our SAF free layer as a function of $V_{bias}$. In addition, $H_C$ represented magnetic anisotropy dramatically increases from ~ 145 Oe to ~ 900 Oe in the FePd SAF MTJs when the $V_{bias}$ sweeps from -0.75 V to +0.75 V, as shown in Fig. S4C, which is more than one order of magnitude larger than that of CFB/MgO/CFB p-MTJs (*14,29*). The VCMA coefficient ($\xi_{VCMA}$) is calculated to be ~ 117 fJ/Vm by fitting the switching field distribution (SFD) from these minor *R-H* loops using the Kurkijärvi-Fulton-Dunkelberger equation (*29*). The large $H_C$ variation and $\xi_{VCMA}$ value are attributed to the anisotropy change of the CFB layer (see Fig. S4D) as well as the modification of the exchange coupling strength between the FePd and CFB layers via E-field.

To further investigate the VCEC switching of p-MTJs, we performed energy calculations. The interplay between VCMA and VCEC can be elucidated from an energetic consideration by extending the Stoner-Wohlfarth model to include VCMA effects. The total energy of MTJ is $E(\theta, V_{bias}) = K_{u,eff}(V_{bias})\sin^2\theta + T_{\perp,IEC}(V_{bias})\cos\theta$, where $T_{\perp,IEC}(V_{bias})$ represents the IEC torque, and $K_{u,eff}(V_{bias})$ is the effective magnetic anisotropy expressed as $K_{u,eff}(V_{bias}) = K_{u,eff}(0) - \xi(V_{bias}/t_{MgO}^2)$ ($t_{MgO}$ is the thickness of MgO barrier) (*29*). The IEC torque possesses a field-like torque behavior (*32-33*). For the MTJ stack with a single ferromagnetic free layer, the IEC torque as a function of the voltage can be defined as $T_{\perp,IEC}(V_{bias}) = a + bV_{bias}^2$ (*33,34*), which shows a quadratic dependence on $V_{bias}$, where *a* and *b* are two numerical parameters describing the magnitude of equilibrium IEC and its voltage modulation, respectively. This quadratic IEC torque cannot lead to a bidirectional magnetization switching in MTJs (*33,34*), as plotted in Fig. 3C. For the MTJ stack with a SAF free layer, e.g. our MTJ stack (FePd/Ta/CFB/MgO/CFB), the IEC torque is a linear function of $V_{bias}$ and, thus, can be described as $T_{\perp,\text{IEC}}(V_{bias}) = a + bV_{bias}$ based on density-functional theory (DFT) calculations shown below. As shown in Fig. 3D, bidirectional magnetization switching is possible by reversing the $V_{\text{bias}}$ polarity. These results indicate that, in an ideal situation where the STT contribution is negligible, as one reverses the direction of the applied E-field, the bidirectional magnetization switching is only possible in p-MTJs having $T_{\perp,\text{IEC}}(V_{bias})$ with a linear voltage dependence. In this simulation, $K_{u,\text{eff}}(0) = 0.51$ Merg/cm$^3$, $\xi = \sim 117$ fJ/Vm and $t_{\text{MgO}} = 2$ nm are obtained from our experimental results, and $a \sim 0.01$ erg/cm$^2$ and $b \sim 0.09$ erg/Vcm$^2$ are obtained from the DFT calculation and Bruno's model.

Figure 4A (top panel) illustrates the SAF structure [FePd/Ta/Co$_2$Fe$_6$ (CoFe)/MgO] from our p-MTJ stack. This SAF structure is employed to understand the VCEC effect through DFT calculations and a quantum interference model proposed by Bruno (see Section S3 in SI) (*23,35*). A schematic of the band diagram is presented in Fig. 3A (bottom panel) emphasizing the linear voltage drop across the MgO tunnel barrier. We defined the reflectivity coefficients at the three interfaces (MgO/CoFe, CoFe/Ta, and Ta/FePd) as $r_A^{\uparrow(\downarrow)}$, $r_B^{\uparrow(\downarrow)}$, and $r_C^{\uparrow(\downarrow)}$ for spin-up and spin-down electrons, respectively. First, the IEC of this SAF structure under different E-fields is investigated by extracting the energy difference $\Delta E = E_{\text{FM}} - E_{\text{AFM}}$ between the FM and AFM exchange coupling configurations. As shown in Fig. 4B, the SAF structure has an energetic preference for the AFM coupling ($\Delta E > 0$) without an E-field, which is consistent with our experimental observation. When applying an E-field ranging from -1.0 V/nm to +1.0 V/nm, a clear transition between the FM and AFM exchange coupling configurations emerges. By analyzing the results, we found that E-field can induce the energetic shifts of the *d*-orbitals and change the magnetic moment of the CoFe layer (see Fig. S6A). Most importantly, the calculated results reveal that there is an intrinsic E-field (intrinsic dipole field) (*11,36*), as shown in Fig. 4C.

This intrinsic E-field originates from the interaction between CoFe and MgO at the interface (see Fig. S6B) and is verified by the change in magnetic moment between interfacial and bulk Fe atoms (see Figs. S7 and S8). Next, we further comprehend the physical mechanism of the VCEC switching. We developed a theoretical scheme based on the quantum interference model proposed by Bruno (*23,35*), as sketched in Fig. 4A (bottom panel). In this model, the IEC between two FM layers depends on the spin-dependent reflection coefficients at the interfaces and can be obtained through Eq. (1):

$$J_{ex} = -\frac{1}{4\pi^3} Im \int d^2k \int_{-\infty}^{+\infty} dE \frac{2\Delta r_A \Delta r_B e^{2ikD}}{1 - 2\bar{r}_A \bar{r}_B e^{2ikD} + (\bar{r}_A^2 - \Delta r_A^2)(\bar{r}_B^2 - \Delta r_B^2) e^{4ikD}} \quad (1)$$

where $\Delta r_{A(B)} = (r_{A(B)}^\uparrow - r_{A(B)}^\downarrow)/2$ and $\bar{r}_{A(B)} = (r_{A(B)}^\uparrow - r_{A(B)}^\downarrow)/2$ are the spin-asymmetry and spin-average of the reflection coefficients. The reflection coefficient is computed exactly by solving the tunneling problem for a trapezoidal barrier. The voltage-dependent reflection coefficient $r_A^{\uparrow(\downarrow)}(V_{bias})$ can be obtained as below (See details in Section S3 in SI):

$$r_A^{\uparrow(\downarrow)}(V_{bias}) = \frac{[ik_B^{\uparrow(\downarrow)} Ai(z_0) - A'i(z_0)][ik_A^{\uparrow(\downarrow)} Bi(z_d) - B'i(z_d)] - [ik_A^{\uparrow(\downarrow)} Ai(z_d) - A'i(z_d)][ik_B^{\uparrow(\downarrow)} Bi(z_0) - B'i(z_0)]}{[ik_B^{\uparrow(\downarrow)} Ai(z_0) + A'i(z_0)][ik_A^{\uparrow(\downarrow)} Bi(z_d) - B'i(z_d)] - [ik_A^{\uparrow(\downarrow)} Ai(z_d) - A'i(z_d)][ik_B^{\uparrow(\downarrow)} Bi(z_0) + B'i(z_0)]} \quad (2)$$

In our simulation configuration, since the CoFe layer was surrounded by the different materials (the MgO tunnel barrier above and the Ta layer below), we expected $r_A^{\uparrow(\downarrow)} \neq r_B^{\uparrow(\downarrow)}$ such that the CoFe layer acts as an asymmetric Fabry-Perot cavity, where the net reflection coefficient that enters in Bruno's expression is

$$r_{B(net)}^{\uparrow(\downarrow)} = \frac{r_{B\infty}^{\uparrow(\downarrow)} + r_A^{\uparrow(\downarrow)} e^{2ik_B^{\uparrow(\downarrow)} t_B}}{1 + r_{B\infty}^{\uparrow(\downarrow)} r_A^{\uparrow(\downarrow)} e^{2ik_B^{\uparrow(\downarrow)} t_B}} \quad (3)$$

Here, $t_B$ is the thickness of the CoFe layer and $r_{B\infty}^{\uparrow(\downarrow)} = (k_F - k_B^{\uparrow(\downarrow)})/(k_F + k_B^{\uparrow(\downarrow)})$ is the reflection coefficient at the CoFe/Ta interface if the corresponding FM layer is infinite. When applying E-field, the reflection coefficient at the MgO/CoFe interface will be changed, leading to the voltage-dependent $J_{ex}(V_{bias})$ via $r_A^{\uparrow(\downarrow)}(V_{bias})$ in Eq. (2) and modifying the reflection coefficient $r_B^{\uparrow(\downarrow)}$ based on Eq. (3), which translates into a voltage dependence of the IEC between CoFe and FePd layers. The spin-dependent reflection coefficient $r_A^{\uparrow(\downarrow)}$ at the MgO/CoFe interface can be defined as $r_A^{\uparrow(\downarrow)} = \exp(i\Phi^{\uparrow(\downarrow)})$, where $\Phi^{\uparrow(\downarrow)}$ represents the voltage-dependent phase factor of the reflected electron wave functions (see Section S3 in SI). This results in the reflection phase, as shown in Fig. 4D. The voltage-dependent phase factor at the CoFe/MgO interface presents a linear dependence on the $V_{bias}$. The phase modulation of ~ 0.029 rads/V/nm and ~ 0.031 rads/V/nm for the spin-up and spin-down electrons are obtained, respectively. Furthermore, the change of the reflection coefficient at the MgO/CoFe interface (see Fig. 4A) can lead to a voltage-dependent IEC energy $\Delta E = 2a^2 J_{ex}$ (35) between the CoFe and FePd layers, where a ~ 2.92 Å is the lattice constant of the configuration obtained from our DFT calculations and $J_{ex}$ is the strength of IEC simulated from Bruno's formula Eq. (1). The evolution of IEC energy $\Delta E$ as a function of E-field is summarized in Fig. 4E, which shows the sign change of the IEC between FePd and CoFe layers, consistent with the DFT calculations. In these calculations, a built-in E-field is considered, as indicated from our DFT calculations. Additionally, we verified that the phase modulation (VCEC effect) is suppressed when the MgO tunnel barrier gets thicker and becomes vanishingly small in the limit of a semi-infinite barrier, where the total phase for the spin up and spin down electrons is -π, as expected.

**DISCUSSION**

We experimentally demonstrated and theoretically confirmed that the bipolar E-field switching of the p-MTJ devices with a SAF free layer. The E-field can modulate the sign of the IEC of the SAF structure and also induce a linear IEC torque, which can switch the p-MTJ devices bidirectionally. The p-MTJ devices switched by the VCEC show an order of magnitude lower energy per write operation compared to that reported STT-switched MTJ devices, and 5.5X higher densities than that for SOT-switched MTJ devices is possible, based on a benchmarking simulation for a 7-nm node architecture. By further optimizing the spacer, ferromagnetic layer, and MTJ stack, VCEC-MTJ could provide many functionalities for realizing the ultralow-energy consumption and ultrafast speed memory and logic applications.

## MATERIALS AND METHODS

**Sample preparation**

The FePd SAF p-MTJ structures studied in this work were prepared on single crystal (001) MgO substrates by magnetron sputtering under an ultrahigh vacuum (base pressure < $5.0 \times 10^{-8}$ Torr) under the same conditions as in our previous work (*37*). The FePd (3 nm)/Ru (1.1 nm)/FePd (3 nm) perpendicular SAF stack was prepared with a Cr (15 nm)/Pt (4 nm) seed layer, keeping the substrate temperature at 350 $^o$C. The rest of the layers of the FePd SAF p-MTJ structures with a stack of Ta (0.8)/CFB (1.3)/MgO (2.0, 2.3)/CFB (1.3)/Ta (0.7)/[Pd (0.7)/Co (0.3)]$_4$/Pd (5)/capping layer (where the numbers in parentheses are thicknesses in nanometers) were grown after the substrate was cooled to room temperature. The 15-nm Pt capping layer was deposited for the conductive atomic force microscope (C-AFM) testing. Before device patterning, the FePd SAF p-MTJ stacks were annealed at 350 $^o$C using rapid thermal annealing (RTA). Then the FePd SAF p-MTJ stacks were patterned into nano-pillars by e-beam lithography and Ar ion milling.

**p-MTJ device testing**

The spin transport properties were tested by using C-AFM at room temperature for 100-nm diameter FePd SAF p-MTJs. The C-AFM was an RHK UHV 350 with an R9 controller operating in contact mode. Si-doped AFM probe tips (Arrow-FM nanoworld) were made conducting by sputtering 200 nm of Pt onto a Ta adhesion layer (*29*). A Pt-coated AFM tip was used to make direct electrical contact with the top of FePd SAF p-MTJ pillars. In all the C-AFM measurements, the tip was grounded and $V_{bias}$ (positive or negative) was applied at the bottom electrode. Thus, for a positive $V_{bias}$, the current flows from the bottom to the top, and for a negative $V_{bias}$, the current flows from the top to the bottom of FePd SAF p-MTJs. The E-field can be calculated by dividing by the thickness of the MgO tunnel barrier through $V_{bias}$. During the testing, a sweep rate of 150 Oe/sec was used to measure the *R-H* loops. A sweep rate of 500 mV/sec was used to measure the resistance versus bias voltage (*R-V*$_{bias}$) loops. For resistance versus time (*R-t*) traces, an acquisition rate of 1 MHz was used. A variable out-of-plane magnetic field with $H_{ext}$ up to 1300 Oe was applied (*29*).

**DFT calculation**

The IEC of the SAF structure as a function of E-field was calculated using first-principles methods based on density functional theory (DFT). As implemented in the Vienna *ab initio* simulation package (VASP) code (*38*), the generalized gradient approximation (GGA) exchange-correlation potentials plus the projector augmented wave (PAW) method for the electron-ion interaction was used (*39*). Based on the experimental structure, we constructed a SAF structure with the stack of MgO (6)/Co$_2$Fe$_6$ (8)/Ta (6)/FePd (11), as shown in Fig. 3A in the main text, to study IEC between Co$_2$Fe$_6$ and FePd layers through a Ta spacer. The numbers indicate the atomic thickness for each layer, which corresponds to a thickness of 10.7, 8.5, 10.8, 15.7 Å for MgO, Co$_2$Fe$_6$, Ta, and FePd, respectively. This Co$_2$Fe$_6$ composition was used because it most closely reflects the stoichiometry after annealing the experimental sample to crystallize the MgO tunnel barrier (*40*). All self-consistent calculations were performed with a plane-wave cutoff of 400 eV. The geometric optimizations were carried out without any constraint until the force on each atom is less than 0.02 eV/ Å and the change of total energy per cell is smaller than $10^{-4}$ eV. The Brillouin zone k-point sampling was set with a 9 × 9 × 1 Γ-centered Monkhorst-Pack grids. A vacuum layer thicker than 15 Å was applied along the *z*-direction to eliminate the interaction between slabs. The lattice constant and the atomic arrangement of the SAF structure were first relaxed without considering spin. The relaxed structure with a lattice constant of 2.92 Å was then applied to study the magnetic interaction between the Co$_2$Fe$_6$ and FePd FM layers. Spin-orbit coupling effects were considered to study the magnetic anisotropy. The IEC was calculated by comparing the energy difference between FM and AFM couplings aligned $Co_2Fe_6$ and FePd layers. An electric field is applied along the *z*-direction with the dipole corrections performed to avoid interactions between the periodically repeated images.

**Acknowledgments:** This work was supported in part by C-SPIN, one of six centers of STARnet, and is currently supported by ASCENT, one of six centers of JUMP, a Semiconductor Research Corporation program that is sponsored by MARCO and DARPA. Portions of this work were conducted in the Minnesota Nano Center, which is supported by the National Science Foundation through the National Nano Coordinated Infrastructure Network (NNCI) under Award Number ECCS-1542202. We would like to thank Dr. Paul M. Haney from the National Institute of Standards and Technology (NIST) for valuable discussions and suggestions, and Prof. Paul M. Voyles form University of Wisconsin-Madison for providing the STEM image.

**Author contributions:** J.P.W initialized and coordinated the project. D.L.Z. and J.P.W. conceived the experiments. D.L.Z. designed all the samples and prepared the MTJ stacks and performed the magnetic characterization. Z.Z., D.L.Z. and D.L. patterned nano-sized MTJs and carried out the spin-transport measurement with Y.L.. M.B. and S.A.M. fabricated the sub-100 nm MTJs and did switching measurements using the C-AFM setup. D.L.Z. and J.P.W. proposed the bipolar E-field included magnetization switching mechanism. W.J., D.P.S., and T.L. did the first-principles and theoretical modeling simulations. Y.C.L. and A.N. did the benchmarking calculations. P.S. helped analyze the *R-H* switching data. D.L.Z. led the manuscript writing with a portion of the materials provided by M.B., W.J., D.P.S, and Y.C.L. All the authors discussed the results and commented on the manuscript.

**Competing interests:** Authors declare no competing interests.

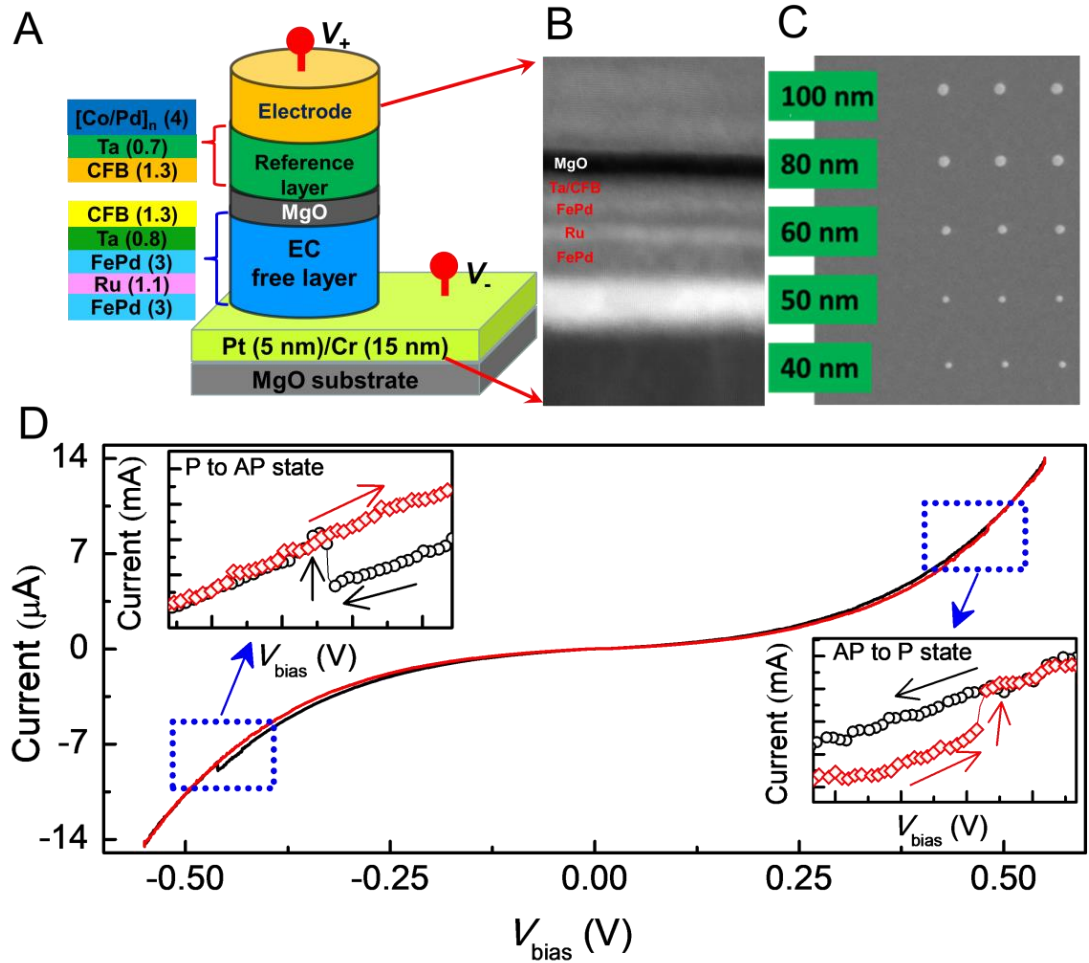

**Fig. 1. Bipolar E-field switching in p-MTJs with ultralow current density.** (**A**) Schematic of the FePd SAF p-MTJ stack. Two FePd layers couple antiferromagnetically through a Ru spacer, then couple with CFB via a Ta spacer, forming the bottom SAF free layer. The CFB/Ta/[Co/Pd]$_n$ stack is as the top reference layer. (**B**), (**C**) The cross-sectional STEM image and the SEM image of FePd SAF MTJ, respectively. (**D**) Current vs. bias voltage ($I$-$V_{bias}$) curve for a 100-nm FePd SAF p-MTJ devices measured by C-AFM without applying $H_{ext}$. Sharp magnetization switching with the parallel (P)-to-antiparallel (AP) state and AP-to-P state are observed by applying a negative $V_{bias}$ ~ -0.45 V and a positive $V_{bias}$ ~ +0.48 V with the switching current density ~ $1.1 \times 10^5$ A/cm$^2$, respectively.

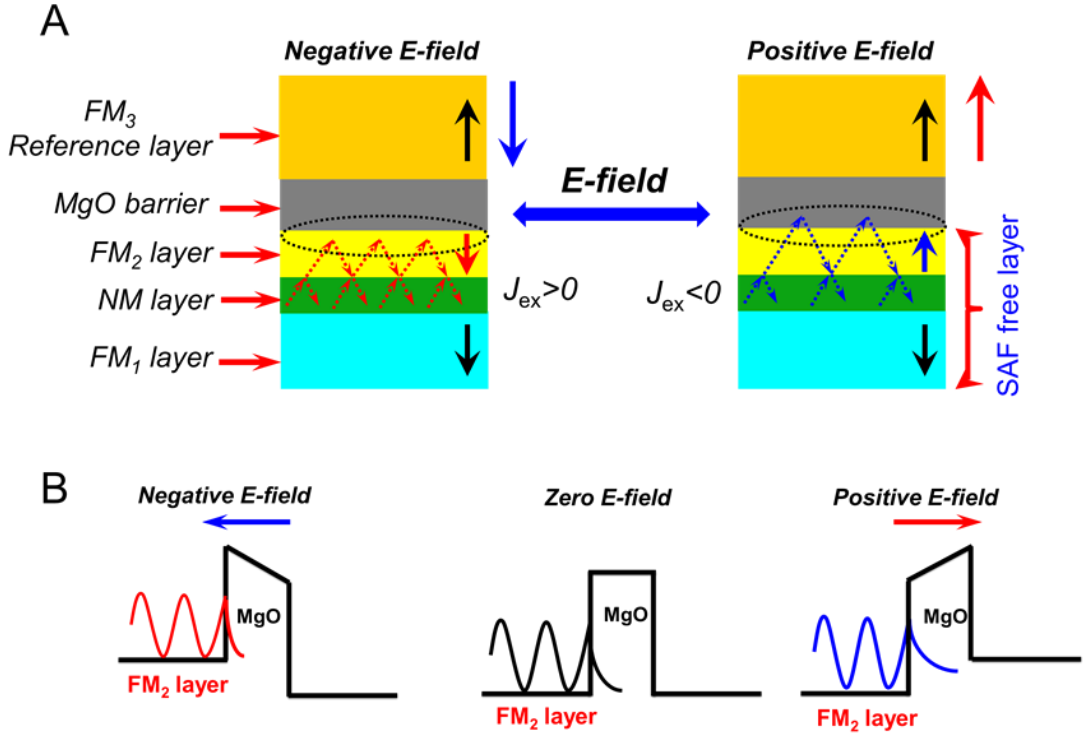

**Fig. 2. Switching mechanism of VCEC-MTJs.** (**A**) The IEC between $FM_1$ and $FM_2$ layers of the SAF free layer can be altered the FM-AFM exchange coupling transition under the applied E-field. In this case, the $FM_2$ and $FM_3$ reference layers will form high or low resistance states. (**B**) The VCEC originates from changes in the reflectivity of the NM/$FM_2$ interface influenced by the $FM_2$/MgO interface. The electron wave functions can penetrate the MgO tunnel barrier, modulating the reflection phases of two-electron spins (spin-up and spin-down) and the penetration length by the applied E-field, then inducing a FM-AFM exchange coupling transition.

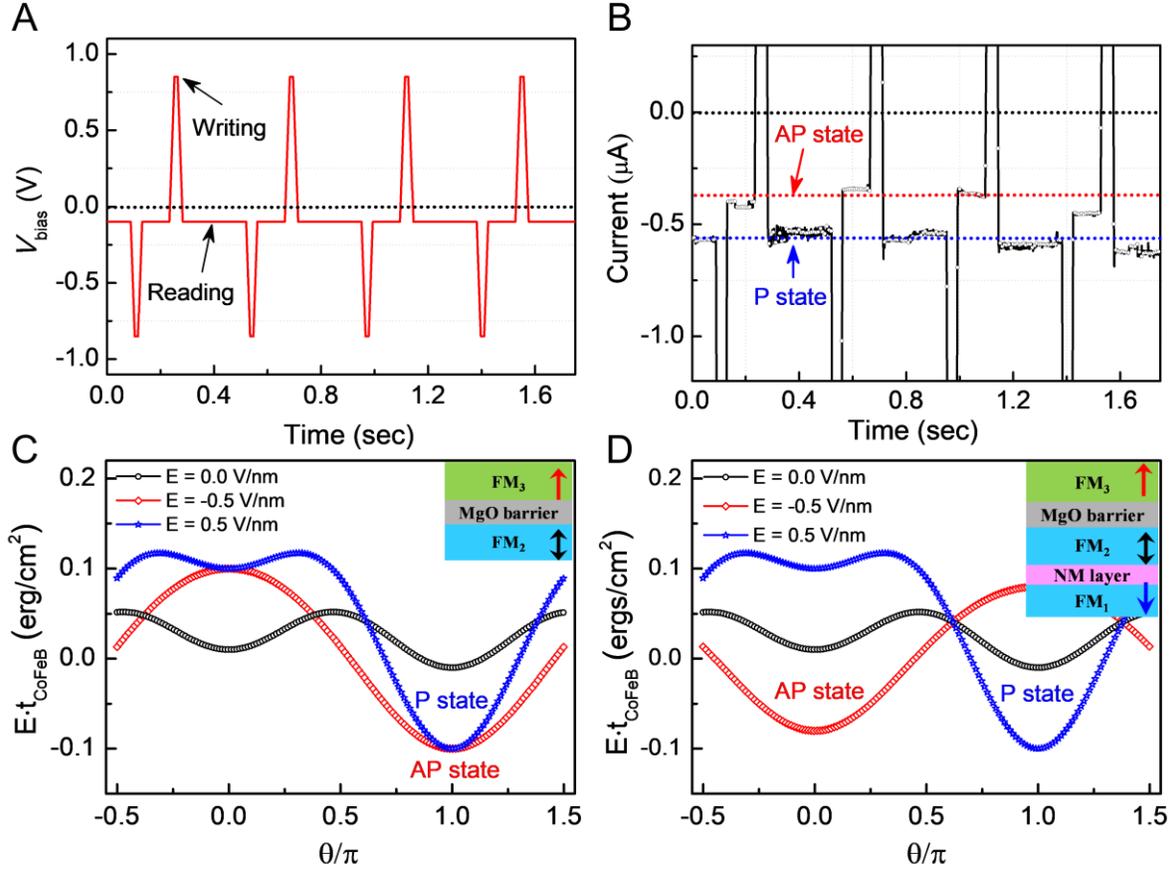

**Fig. 3. Robustness of bipolar E-field switching in p-MTJs.** (**A**) Negative and positive 0.85 V voltage pulses were applied for a *write* operation, and a negative 0.1 V was used to *read* the current levels to check the device state (parallel or antiparallel states). (**B**) The current vs. time trace measured for the same device without $H_{ext}$. Clear bidirectional magnetization switching between parallel and antiparallel states with the bipolar $V_{bias}$. (**C**) The IEC torque in the p-MTJ stack with a single free layer can be defined as $T_{\perp,IEC}(V) = a + bV^2$, which shows the quadratic behavior depending on $V_{bias}$ and cannot switch the magnetization of p-MTJs. The magnetization of p-MTJ prefers the antiparallel configuration with the bipolar $V_{bias}$. (**D**) For the p-MTJ stack with a SAF free layer, e.g. our MTJ stack (FePd/Ta/CFB/MgO/CFB), the IEC shows the linear trend as a function of $V_{bias}$ (shown in Fig. 4B) with a description as $T_{\perp,IEC}(V) = a + bV$. This IEC torque can bidirectionally switch the magnetization of p-MTJs with the bipolar $V_{bias}$, like the functionality of a STT effect.

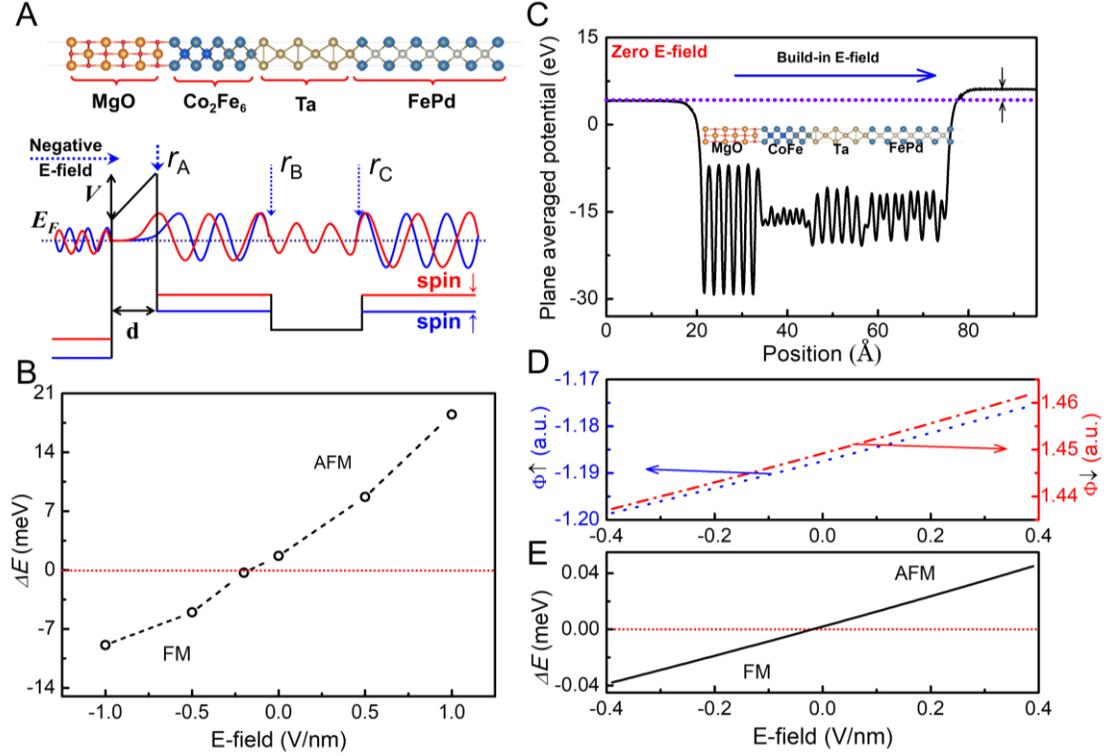

**Fig. 4. Voltage-controlled exchange coupling (VCEC). (A) top,** Schematic of the simulation structure with a SAF stack of FePd/Ta/Co$_2$Fe$_6$ (CoFe)/MgO (top panel) which is comparable to the experimental SAF structure. (**A**) **bottom**, Schematic of the energy diagram of the SAF structure (bottom panel), where we define the interfacial reflection coefficient $r_A$, $r_B$, and $r_C$ corresponding to the three interfaces: CoFe/MgO, Ta/CoFe, and FePd/Ta, respectively. The ferromagnetic layers are characterized by an exchange splitting, which separates the majority and minority spin bands. The voltage is effectively dropped inside the MgO tunnel barrier, changing the reflection coefficient of electron wave functions at this interface, and modulating the IEC. (**B**) $\Delta E$ (IEC) vs. E-field for the FePd/Ta/CoFe/MgO SAF structure calculated by DFT. (**C**) An intrinsic built-in voltage potential generates the built-in E-field due to interaction at the MgO/CoFe interface, which plays a very critical role for the E-field induced AFM-FM exchange coupling transition. By applying the E-field, the voltage drop inside the MgO tunnel barrier modulates the effective electron transversal path length. This results in the reflection phase, as shown in (**D**), which in turn modifies the electronic density of states in the spacer layer through quantum interferences leading to a controllable transition between FM and AFM exchange couplings. (**E**) IEC as a function of the applied voltage calculated using Bruno's model.